\documentclass[twocolumn]{emulateapj}
\usepackage{graphicx}
\usepackage{epsfig}
\usepackage{amsmath}

\usepackage{color}

\shorttitle{Neutron-Star Parameters}
\shortauthors{Baub\"ock et al.}

\begin{document}
\title{Relations Between Neutron-Star Parameters in the Hartle-Thorne
  Approximation} \author{Michi Baub\"ock\altaffilmark{1}, Emanuele
  Berti\altaffilmark{2,3}, Dimitrios Psaltis\altaffilmark{1,4}, and
  Feryal \"Ozel\altaffilmark{1,4,5}}

\altaffiltext{1}{Astronomy Department, University of Arizona, 933
  North Cherry Avenue, Tucson, AZ 85721, USA}

\altaffiltext{2}{Department of Physics and Astronomy,
The University of Mississippi,
University, MS 38677, USA}

\altaffiltext{3}{Theoretical Astrophysics 350-17, California Institute of Technology, Pasadena, CA 91125, USA}

\altaffiltext{4}{Institute for Theory and Computation,
  Harvard-Smithsonian Center for Astrophysics, 60 Garden Street,
  Cambridge, MA 02138, USA}

\altaffiltext{5}{Radcliffe Institute for Advanced Study, Harvard
  University, 8 Garden Street, Cambridge, MA 02138, USA}

\email{email: mbaubock@email.arizona.edu, berti@phy.olemiss.edu,
  dpsaltis@email.arizona.edu, fozel@email.arizona.edu}

\begin{abstract} 

Using stellar structure calculations in the Hartle-Thorne
approximation, we derive analytic expressions connecting the
ellipticity of the stellar surface to the compactness, the spin
angular momentum, and the quadrupole moment of the spacetime. We also
obtain empirical relations between the compactness, the spin angular
momentum, and the spacetime quadrupole. Our formulae reproduce the
results of numerical calculations to within a few percent and help
reduce the number of parameters necessary to model the observational
appearance of moderately spinning neutron stars. This is sufficient
for comparing theoretical spectroscopic and timing models to
observations that aim to measure the masses and radii of neutron
stars and to determine the equation of state prevailing in their
interiors.

\end{abstract}

\section{Introduction}

Measuring the masses and radii of neutron stars provides one of the
most stringent tests of our understanding of the properties of matter
under extreme conditions. Several methods of measuring these
properties involve analyzing the emission originating from the stellar
surface. In order to correctly interpret this surface emission, it is
necessary to understand the strong-field gravitational lensing
experienced by the photons when traveling through the curved spacetime
in the vicinity of a neutron star.

To date, considerable effort has been expended to accurately measure
neutron-star radii, primarily through the spectroscopic observations
of quiescent neutron stars (e.g., Heinke et al.\ 2006; Webb \& Barret
2007; Guillot et al.\ 2011) and X-ray bursters (e.g., \"Ozel et
al.\ 2009: see \"Ozel 2013 for a recent review). In the near future,
X-ray missions such as NICER and LOFT as well as gravitational-wave
detectors such as Advanced LIGO will allow even more precise
measurements of various neutron star properties.

Many of the primary targets for future measurements have moderate
spins ($\sim$300--800~Hz; e.g., Galloway et al.\ 2008; Bogdanov et
al.\ 2008). At these frequencies, gravitational effects depend not
only on the mass and radius, but also on other parameters, such as the
quadrupole moment of the neutron star and the oblateness of its
surface (Morsink et al.\ 2007; Baub\"ock et al.\ 2013). Exploiting the
upcoming high-quality observations and measuring the masses and radii
of neutron stars at the accuracy necessary to constrain their equation
of state requires taking these non-negligible effects into account.

For the moderate spin frequencies of weakly magnetic neutron stars,
the Hartle-Thorne metric provides an accurate approximation to their
spacetime (Hartle \& Thorne 1968). In this regime, the appearance of
a neutron star as measured by an observer at spatial infinity depends on seven macroscopic parameters:
the mass $M$, the equatorial radius $R_{\rm eq}$, the spin frequency
$f$, the inclination of the rotational pole with respect to the
observer $\theta_0$, the angular momentum $J$, the quadrupole moment
$Q$, and the eccentricity of the surface $e_s$. Three of these
parameters ($f$, $\theta_0$, and, e.g., $M$) are independent of the equation of state for any given
observed source. The remaining four parameters ($R_{\rm eq}$, $J$,
$Q$, and $e_s$) are uniquely determined by the equation of state,
given a neutron star mass and spin frequency. Therefore, it is any of these
four parameters that need to be measured observationally, in addition to the mass and spin frequency, in
order for the underlying equation of state to be constrained.

Even though measuring only one of the four dependent parameters with
high precision would be sufficient, typical observables depend on all
seven parameters in a complex manner.  It is unlikely that
spectroscopic or timing observations in the near future will be
accurate enough to allow for independent measurements of all of these
parameters in individual neutron stars. In order to make progress, we
can reduce the dimensionality of the problem by using approximate
relations that connect quantities that are higher order in spin
frequency (such as the spin angular momentum $J$, the spacetime
quadrupole $Q$, and the ellipticity of the stellar surface $e_s$) to
ones that are of lower order (such as the mass $M$ and equatorial
radius $R_{\rm eq}$). In this way, observable phenomena from a
moderately spinning neutron star can be calculated based only on its
mass and equatorial radius, given a spin frequency and an observer's
inclination.

This reduction of the parameter space by means of approximate
relations allows for the properties of dense matter to be constrained
only if the relations themselves do not depend strongly on the details
of the equation of state. Andersson \& Kokkotas (1998) modeled pulsation 
modes of neutron stars and showed that the relations between several 
parameters of interest have a significant dependence on the equation of state. However, 
given the constraints imposed on the equation of state of dense matter by 
recent observations (e.g., Demorest et al.\ 2010; Antoniadis et al.\ 2013),
we show that it is possible to find relations between the parameters 
described above that are valid over the astrophysically relevant parameter
range and for a variety of equations of state.

Several authors to date have explored such approximate relations in
different contexts. Ravenhall \& Pethick (1994) and Lattimer \&
Prakash (2001) provide empirical formulae for the moments of inertia
and binding energies of slowly spinning neutron stars as a function of
their masses and radii. Morsink et al.\ (2007) obtained empirical
formulae that connect the ellipticity of the surfaces of spinning
neutron stars to their masses, equatorial radii, and spin
frequencies. More recently Urbanec et al.\ (2013) modeled the angular
momenta and quadrupole moments of both neutron stars and strange
stars, showing that different relations exist for these two classes of
objects. Finally, Yagi \& Yunes (2013a) found relations between the
moment of inertia, the quadrupole moment, and the tidal Love number
that are highly accurate for several equations of state.

In this paper, we model the properties of moderately spinning neutron
stars in the Hartle-Thorne approximation, which is adequate for spin
frequencies $\lesssim 800$~Hz. We derive an analytic expression
connecting the ellipticity of the stellar surface to the compactness,
the spin angular momentum, and the spacetime quadrupole. We also
obtain empirical relations between the compactness, the spin angular
momentum, and the spacetime quadrupole similar to those found in Lattimer 
\& Prakash (2001) and Yagi \& Yunes (2013a). These relations allow us 
to fully determine the parameters of a neutron star given a measurement 
of its mass, radius, and spin frequency. We demonstrate that our
formulae reproduce the results of numerical calculations of
neutron-star spacetimes to within a few percent. This is sufficient
for comparing theoretical spectroscopic and timing models to
observations that aim to measure the masses and radii of neutron stars
and to determine the high-density equation of state prevailing in
their interiors.

\section{Numerical models in the Hartle-Thorne approximation}

The Hartle-Thorne metric is based on a slow-rotation expansion. If the
expansion is truncated at second order in the spin frequency, the
spacetime exterior to a rotating object can be described by three
parameters: the total mass, the angular momentum, and the quadrupole
moment of a neutron star. Observationally, the appearance also depends on
the geometry of its surface, i.e., on its equatorial radius and
ellipticity.

Out of these parameters, we choose the mass $M$ and the equatorial
radius $R_{\rm eq}$ to characterize each neutron star. The other three
parameters depend on the equation of state but are of higher order in
spin frequency and introduce small corrections to most observables
(e.g., Poisson 1998; Morsink et al.\ 2007; Racine 2008; Baub\"ock et
al.\ 2013; Psaltis \& \"Ozel 2013).  Therefore, we aim to find
relations that allow for these parameters to be approximated given a
neutron-star mass and radius, independent of the equation of state.

The angular momentum $J$ of a neutron star is often represented by the
dimensionless spin parameter
\begin{equation}
a \equiv \frac{c J}{G M^2},
\label{eq:a_def}
\end{equation}
which is zero for a non spinning object. Neutron stars typically have a
spin $a \leq 0.7$ for uniform rotation and physically motivated
equations of state \citep{coo94,bs04,ll2011}, but the spin magnitudes
of neutron stars in binaries observable by Advanced LIGO are likely to
be much smaller than this theoretical upper bound
\citep{mos10,bhl12}. The spin periods of isolated neutron stars at
birth should be in the range 10-140~ms \citep{lor01}, or $a\lesssim
0.04$. Accretion from a binary companion can spin up neutron stars but
is unlikely to produce periods less than 1 ms, i.e., $a\lesssim 0.4$
\citep{fer08}. The fastest-spinning observed pulsar has a period of
1.4~ms, ($a\sim 0.3$) \citep{hes06}; the fastest known pulsar in a
neutron star-neutron star system, J0737-3039A, has a period of
22.70~ms ($a \sim 0.02$; Burgay et al.\ 2003).
The spin parameter depends both on the rotational period and on the
moment of inertia of the neutron star, which is determined by the
equation of state.

A spinning neutron star also acquires a nonzero quadrupole moment
$Q$. We characterize the quadrupole moment by the dimensionless
quantity
\begin{equation}
q \equiv -\frac{c^4 Q}{G^2 M^3}.
\label{eq:q_def}
\end{equation}
Laarakkers \& Poisson (1999), \cite{bs04}, and Pappas \& Apostolatos
(2012) computed the quadrupole moment of rapidly spinning neutron
stars for a range of equations of state. They found values of $q$
ranging between 1 and 11 (see also Baub\"ock et al.\ 2012).

Lastly, a spinning neutron star also becomes oblate in shape. In the
Hartle-Thorne approximation, this oblateness is described by
\begin{equation}
R(\theta) = R_0 + \xi_2 P_2(\cos \theta),
\label{eq:legendre}
\end{equation}
where $P_2$ is the second-order Legendre polynomial and $\xi_2$ is a
coefficient depending on the equation of state and spin frequency of
the neutron star. In the non-spinning limit, $\xi_2 = 0$ and
$R(\theta) = R_0$. For moderately spinning neutron stars, there are
two frequently used parameters to characterize the oblate shape: the
eccentricity of the surface,
\begin{equation}
e_s \equiv \sqrt{\left(\frac{R_{\rm eq}}{R_{\rm p}}\right)^2 - 1},
\label{eq:es_def}
\end{equation}
and its ellipticity,
\begin{equation}
\varepsilon_s \equiv 1 - \frac{R_{\rm p}}{R_{\rm eq}},
\label{eq:el_def}
\end{equation}
where $R_{\rm eq}$ is the equatorial radius and $R_{\rm p}$ is the
radius at the pole.

As in Berti et al.\ (2005), we expand the parameters $a$, $q$, $e_s$, and $\varepsilon_s$ to second order
in the spin frequency of the neutron star. Specifically, we define the
parameter
\begin{equation}
\epsilon_0 \equiv \frac{f}{f_0}
\label{eq:eps_def}
\end{equation}
in terms of the characteristic frequency
\begin{equation}
f_0 \equiv \sqrt{\frac{G M_0}{R_0^3}}.
\label{eq:O0_def}
\end{equation}
In this equation, $M_0$ and $R_0$ are the non-spinning mass and radius
of the neutron star. The characteristic frequency $f_0$ corresponds to
the Keplerian orbital period of a test particle at a radius $R_0$
around a mass $M_0$ and thus corresponds roughly to the maximum
frequency a neutron star can be spun up to before breakup.  For spin
frequencies much smaller than this characteristic frequency ($f <
f_0$), $\epsilon_0$ serves as a suitable small parameter about which
we can expand the metric. When $f$ approaches $f_0$, the parameter
$\epsilon_0$ approaches unity, and the Hartle-Thorne approximation is
no longer valid. The spin frequency at which this occurs depends on
$M_0$ and $R_0$ and, therefore, on the equation of state. However, for
most proposed equations of state, this approximation is valid for even
the most rapidly spinning neutron stars observed to date~\citep{ber05}.

For a non-spinning neutron star, the parameter $\epsilon_0$ is equal
to zero, and thus the spin $a$, the quadrupole moment $q$, and the
eccentricity of the surface $e_s$ are all zero, as well. As the spin
frequency increases, corrections to the metric enter at different
orders in $\epsilon_0$. To first order in the spin frequency, the star
acquires a non-zero angular momentum, characterized by the spin
parameter $a$. To the lowest order, we can approximate the spin
parameter as a linear function of spin frequency, i.e.,
\begin{equation}
a = \epsilon_0 a^*,
\label{eq:a_spin}
\end{equation}
where $a^*$ is a constant that depends on the equation of state. To
second order in the spin frequency, the star acquires a quadrupole
moment and an elliptical shape, i.e.,
\begin{equation}
q = \epsilon_0^2 q^*
\label{eq:q_spin}
\end{equation}
and
\begin{equation}
\frac{R_{\rm eq}}{R_{\rm p}} = 1 + \epsilon_0^2 R^*,
\label{eq:Rratio_spin}
\end{equation}
where $q^*$ and $R^*$ are again constants depending on the equation of
state. Substituting Equation (\ref{eq:Rratio_spin}) into
Equations~(\ref{eq:es_def}) and (\ref{eq:el_def}) shows that the
eccentricity of the surface of the neutron star has a first order
dependence on the spin frequency
\begin{equation}
e_s = \epsilon_0 e_s^*,
\label{eq:es_spin2}
\end{equation} 
while the ellipticity has a second-order dependence on the spin frequency, i.e.,
\begin{equation}
\varepsilon_s = \epsilon_0^2 \varepsilon_s^*.
\label{eq:el_spin}
\end{equation}

The above relations all depend on $\epsilon_0$, which in turn depends
on the non-spinning values $M_0$ and $R_0$. For a spinning neutron
star, however, these quantities are not readily measurable. Instead,
observations can only constrain the spinning mass and equatorial
radius, $M$ and $R_{\rm eq}$, respectively. These parameters differ
from their non-spinning values at second order in $\epsilon_0$:
\begin{eqnarray}
M = M_0 + \epsilon_0^2 \delta M^*, \\
R_{\rm eq} = R_0 + \epsilon_0^2 \delta R^*,
\label{eq:dM_dR}
\end{eqnarray}
where $\delta M^*$ and $\delta R^*$ are again constants depending on
the equation of state. Since $M$ and $R_{\rm eq}$ differ from $M_0$
and $R_0$ at second order in $\epsilon_0$, the corrections introduced
to Equations (\ref{eq:a_spin})--(\ref{eq:es_spin2}) by the altered
mass and radius will necessarily enter at third or fourth order in
$\epsilon_0$. Therefore, the lowest-order effects will be
unchanged. For the remainder of this work, we will use the spinning
mass and radius interchangeably with the nonspinning values.

We use the procedure described in Berti et al.\ (2005) to calculate
the values of the parameters described above for several neutron-star
equations of state. For a given equation of state, a central density
and spin frequency uniquely determine the properties of a neutron star
in the Hartle-Thorne approximation. First, we solve the
Tolman-Oppenheimer-Volkoff equations to find the parameters of a
non-spinning star with the same equation of state and central
density. In this non-spinning case, $a$, $q$, and $e_s^*$ are equal to 0. 
Next, we solve the full Hartle-Thorne equations for the perturbative 
quantities, i.e. $q^*$, $a^*$, $R^*$, $e_s^*$, and $\varepsilon_s^*$. 
Once we have found the values of these starred parameters, we can then
use Equations (\ref{eq:a_spin})--(\ref{eq:el_spin}) to determine the parameters 
of a neutron star spinning at any intermediate rate characterized by 
$\epsilon < \epsilon_0$.

\section{Relations Between Spin, Quadrupole, and Compactness}

As in Lattimer \& Prakash (2001) and Yagi \& Yunes (2013a), 
we find that tight empirical relations exist between the spin
parameter $a^*$, the dimensionless quadrupole moment $q^*$, and the
compactness $\zeta = G M_0 / R_0 c^2$ of neutron stars that 
depend very weakly on the assumed equation of state. In addition, we
derive an analytic formula relating these four quantities to the
eccentricity parameter $e_s^*$ and the ellipticity parameter
$\varepsilon_s^*$ of the neutron star surface.

In order to generate our fits, we selected several modern equations of
state. Observations by Demorest et al.\ (2010) of a 1.97~$M_\odot$
neutron star and by Antoniadis et al.\ (2013) of a 2.01~$M_\odot$
neutron star place significant constraints on the properties of dense
matter and strongly disfavor several equations of state. We selected
only equations of state that allow a maximum mass of at least
2.0~$M_\odot$. Following the naming convention of Lattimer \& Prakash
(2001), we chose for our fits the equations of state AP4, ENG, MPA1,
and MS0, which cover a wide range of microphysics assumptions and
calculational procedures.

\begin{figure}
\psfig{file=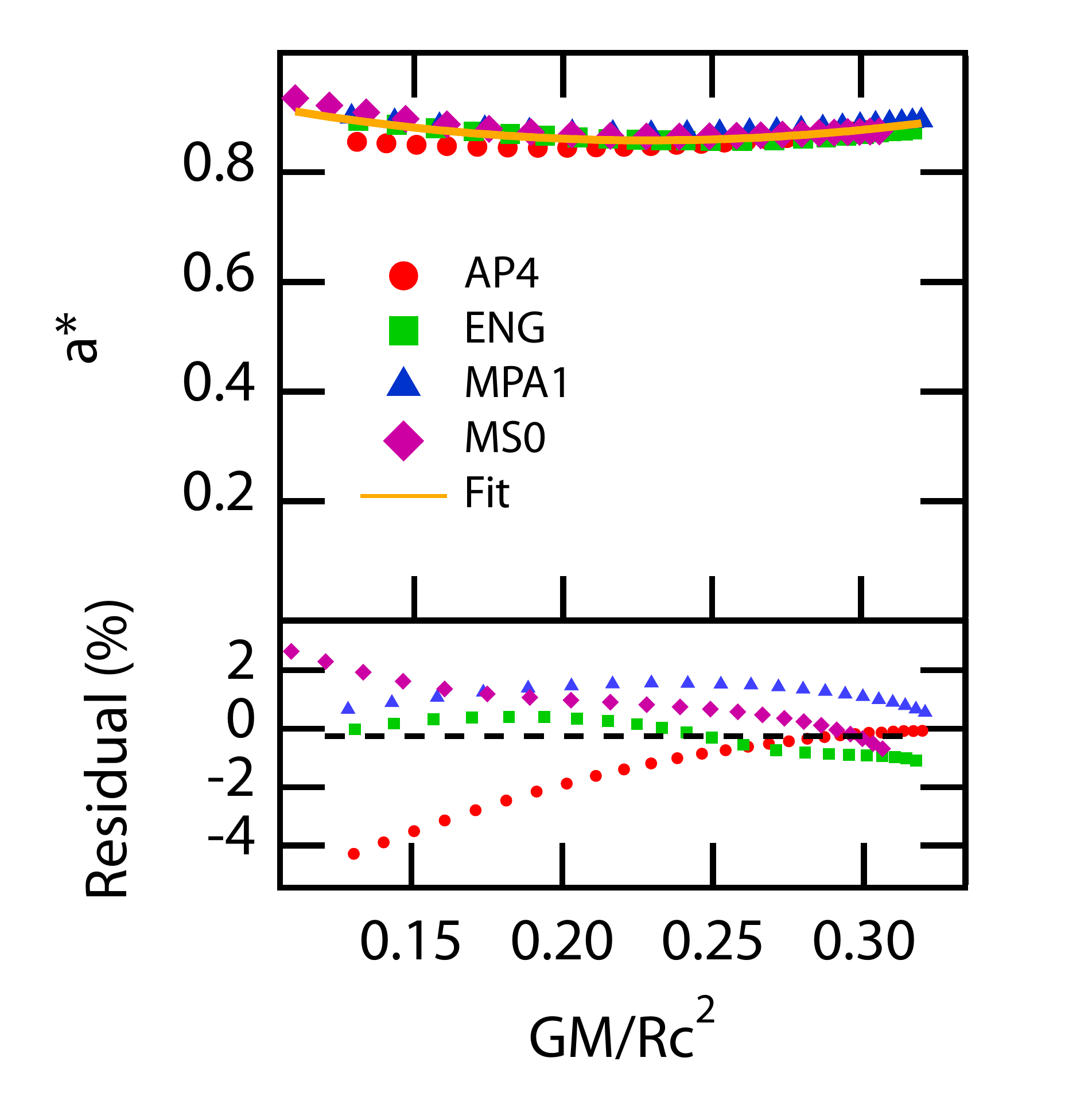, width = 3.5in}
\caption{Empirical fit to the correlation between spin and compactness
  of a neutron star for four equations of state, corresponding to
  equation~(\ref{eq:J_fit}). The lower panel shows the residual in
  percent.}
\label{fig:J_Fit}
\end{figure}

For each equation of state, we use a large number of numerical models
covering the astrophysically relevant range of masses $M>1.0~M_\odot$
(\"Ozel et al 2012). A least-squares polynomial fit of the spin
parameter $a^*$ as a function the compactness yields
\begin{equation}
a^* = 1.1035 - 2.146 \zeta + 4.5756 \zeta^2\,.
\label{eq:J_fit}
\end{equation}
Figure~\ref{fig:J_Fit} shows this fit as a solid line, along with the
results of numerical calculations for four different equations of
state. The lower panel shows the residuals: for all equations of
state, the residuals over the range of masses considered here are less
than 4\%.

Both Lattimer \& Prakash (2001) and Yagi \& Yunes (2013b) found
similar empirical relations between the moment of inertia and the 
compactness. These authors consider a wider range of equations of state 
and neutron star parameters than those included in Figure~\ref{fig:J_Fit}. 
For less compact neutron stars than those shown in Figure~\ref{fig:J_Fit},
i.e., typically those with masses \textless1~$M_\odot$,
different equations of state predict more divergent values for the moment
of inertia (or equivalently the spin parameter $a^*$). For the purpose 
of modeling observations of astrophysical neutron stars, however, the
relation given in Equation~(\ref{eq:J_fit}) is adequate.

\begin{figure}
\psfig{file=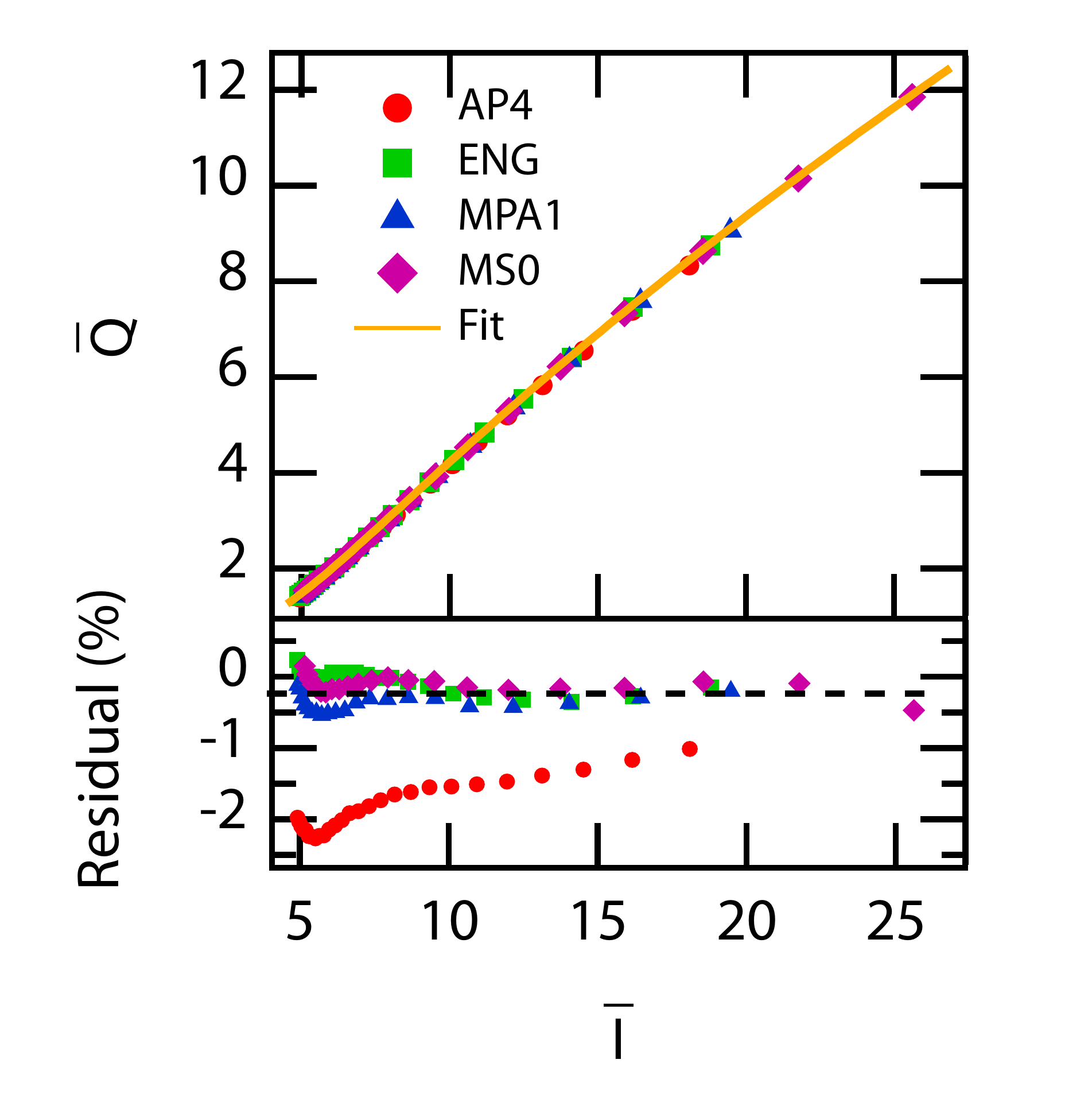, width = 3.5in}
\caption{Empirical fit to the correlation between the dimensionless
  quadrupole moment $\bar{Q}$ and angular momentum $\bar{I}$. The fit
  corresponds to Equation~(\ref{eq:Ibar_Fit}) and is equivalent to
  that proposed in Yagi \& Yunes (2013b). The lower panel shows the
  residual to the fit in percent.}
\label{fig:Q_Fit}
\end{figure}

In order to determine the quadrupole moment, we adopt the relation
proposed by Yagi \& Yunes (2013b). These authors present a relation
between the quadrupole moment and moment of inertia of spinning
neutron stars with a variety of equations of state. They define a
dimensionless quadrupole moment $\bar{Q}$ and moment of inertia
$\bar{I}$ that relate to our $q^*$ and $a^*$ via
\begin{equation}
\overline{Q} \equiv \frac{q^*}{a^{*2}},
\label{eq:Qbar}
\end{equation}
and
\begin{equation}
\overline{I} \equiv a^* \zeta^{-3/2}.
\label{eq:Ibar}
\end{equation}
They then find an empirical expression for $\bar{I}$ as a function of
$\bar{Q}$. Since the inverse of this relation is required for use with
our fit for the spin parameter $a^*$ in Equation~(\ref{eq:J_fit}), we
find instead an analogous fit for $\bar{Q}$ as a function of
$\bar{I}$,
\begin{eqnarray}
\label{eq:Ibar_Fit}
\ln{\overline{Q}} &=& -2.014 + 0.601 \ln{\overline{I}} + 1.10 (\ln{\overline{I}})^2 \\
&-& 0.412 (\ln{\overline{I}})^3 + 0.0459 (\ln{\overline{I}})^4.
\nonumber
\end{eqnarray}
Alternatively, the fit of Yagi \& Yunes (2013b) can be inverted
numerically to obtain an equivalent relation. The above fit is shown
as the solid line in Figure~\ref{fig:Q_Fit} along with numerical
calculations for our chosen equations of state. Again, the residuals
shown in the lower panel are less than 2\% for each considered
equation of state.

\section{Relations for the Elliptical Shape of the Neutron Star Surface}
Given these two empirical fits for the spin parameter and quadrupole
moment, we now find an analytic expression for the eccentricity of the
neutron star surface. Hartle \& Thorne (1968) solved the equations of
stellar structure at second order in spin frequency and showed that the
eccentricity of the neutron star surface measured in flat space is
given by
\begin{equation}
e_s = \sqrt{-3(v_2 - h_2 + \xi_2/R_0)},
\label{eq:HT25c}
\end{equation}
where $v_2$ and $h_2$ are functions of $R_0$ that are second order in
spin, and $\xi_2$ is the parameter defined in
Equation~(\ref{eq:legendre}). Hartle \& Thorne (1968) provide 
the exact forms of the quantities $v_2$ and $h_2$ as functions of the mass, radius, 
angular momentum, and quadrupole moment. Using Equations (\ref{eq:a_def}) 
and (\ref{eq:q_def}), we can reduce Equation~(\ref{eq:HT25c}) to depend 
only on the dimensionless parameters $\zeta$, $a$, $q$, and $\epsilon_0$ 
defined above. Substituting $a^*$ and $q^*$ for $a$ and $q$ via
Equations (\ref{eq:a_spin}) and (\ref{eq:q_spin}) and using the 
definition of $e_s^*$ in Equation (\ref{eq:es_spin2}), we can eliminate
the dependence on spin and find an analytic expression for the eccentricity $e_s^*$,
\begin{multline}
e_s^*(\zeta, a^*, q^*) = \bigg[1 - 4 a^* \zeta^{3/2} \\
+ \frac{15 (a^{*2} - q^*)(3 - 6\zeta + 7\zeta^2)}{8 \zeta^2} + \zeta^2 a^{*2} (3 + 4 \zeta)\\
+ \frac{45}{16 \zeta^2} (q^* - a^{*2})(\zeta - 1)(1 - 2 \zeta + 2 \zeta^2) \ln{(1 - 2\zeta)}\bigg]^{1/2}.
\label{eq:es_analytic}
\end{multline}

\begin{figure}
\psfig{file=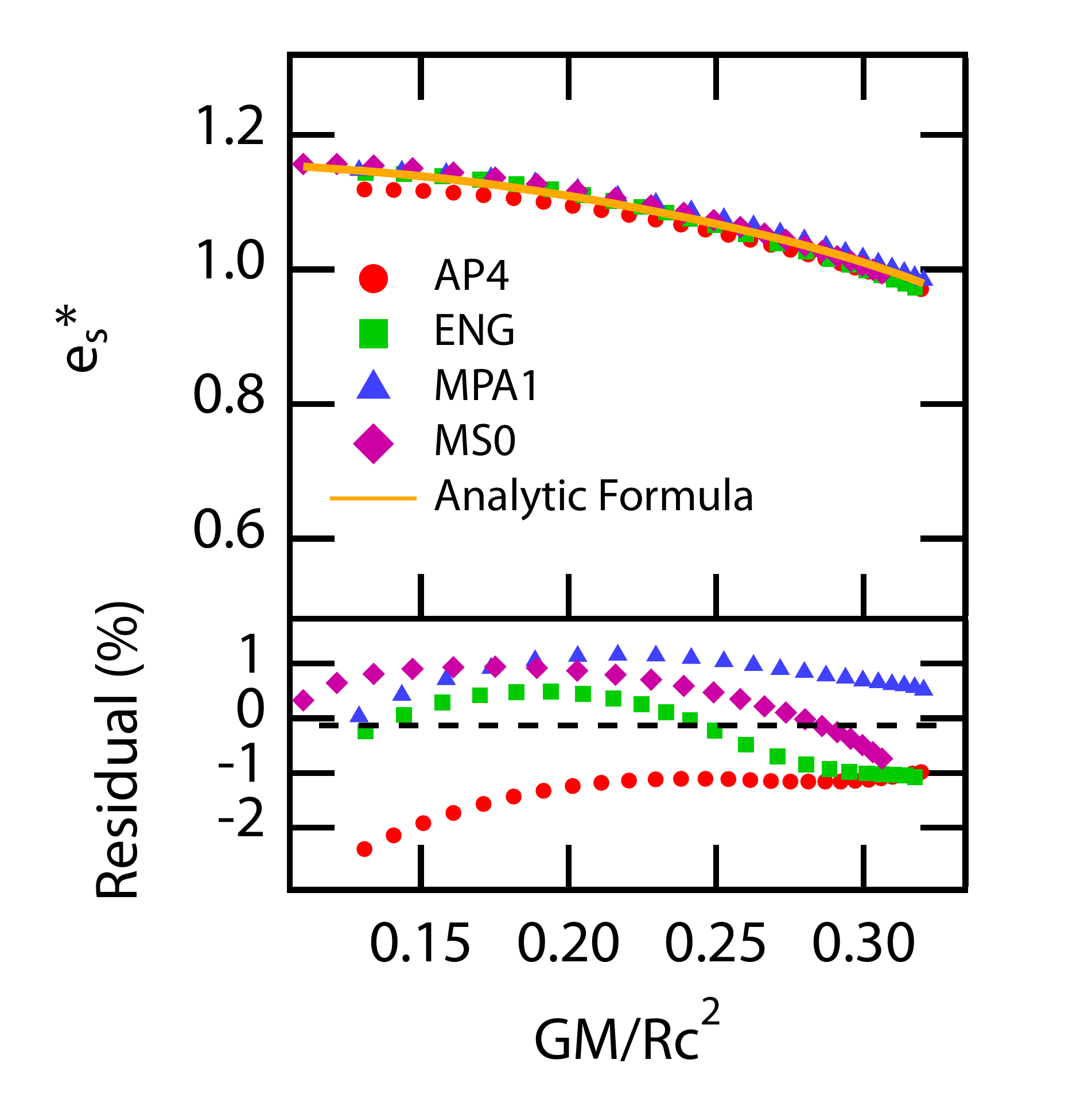, width = 3.5in}
\caption{Analytic expression for the eccentricity of the neutron star
  surface, corresponding to Equation~(\ref{eq:es_analytic}). In order
  to express this relation as a function of a single parameter, we
  have combined Equation~(\ref{eq:es_analytic}) with our empirical
  fits for the quadrupole moment and spin parameter from
  Equations~(\ref{eq:J_fit}) and (\ref{eq:Ibar_Fit}). The lower panel
  shows the residual to the relation in percent. Although the analytic
  relation is exact, the empirical fits between $a^*$, $q^*$, $R$, and
  $M$ introduce some scatter about this relation.}
\label{fig:es_Fit}
\end{figure}

Figure~\ref{fig:es_Fit} shows this expression for the eccentricity as
a function of the compactness, along with numerical calculations from
several equations of state. We have substituted
Equations~(\ref{eq:J_fit}), (\ref{eq:Ibar}) and (\ref{eq:Ibar_Fit})
into Equation~(\ref{eq:es_analytic}) in order to present the relation
as a function of the single parameter $\zeta$. The residuals to this
relation are shown in the lower panel. The residuals are nonzero due
to the empirical nature of the fits between $a^*$, $q^*$, and $\zeta$.

Alternatively, Hartle (1967) gives an expression for the ellipticity
of the neutron star surface in Hartle-Thorne coordinates as
\begin{equation}
\varepsilon_s = - \frac{3}{2 R} \xi_2.
\label{eq:H146c}
\end{equation}
Again, we can reduce this equation to depend only on our dimensionless
parameters and eliminate the spin dependency to find an expression for 
$\varepsilon_s^*$, 

\begin{multline}
\varepsilon_s^*(\zeta, a^*, q^*) = \frac{1}{32 \zeta^3} \Bigg\{2 \zeta \Big[8 \zeta^2 - 32 a^* \zeta^{7/2} \\
 + (a^{*2} -  q^*) (45 - 135  \zeta  + 60 \zeta^2  + 30 \zeta^3 ) + 24 a^{*2} \zeta^4 + 8 a^{*2} \zeta^5\\
  - 48 a^{*2}\zeta^6 \Big] + 45 (a^{*2} - q^*)(1 - 2\zeta)^2 \ln{(1 - 2\zeta)}\Bigg\}. 
\label{eq:el_analytic}
\end{multline}


\section{Applications}

In order to reduce the number of parameters necessary when fitting a
neutron star observation, the following procedure can be used. For a
given value of $M$, $R_{\rm eq}$, and $f$, one can calculate the
parameter $\epsilon_0$ via Equations~(\ref{eq:eps_def}) and
(\ref{eq:O0_def}) as
\begin{equation}
\epsilon_0 = f \left(\frac{G M}{R_{\rm eq}^3}\right)^{-1/2}.
\label{eq:rec_eps0}
\end{equation}
The spin parameter $a$ can then be found from Equation~(\ref{eq:J_fit}), 
\begin{equation}
a = \epsilon_0 \left[1.1035 - 2.146 \zeta + 4.5756 \zeta^2\right].
\label{eq:rec_a}
\end{equation}
The fit between moment of inertia and the quadrupole moment can then
be used to write
\begin{multline}
q =a^2 \exp\Bigg[-2.014 + 0.601 \ln{\left(\frac{a}{\epsilon_0} \zeta^{-3/2}\right)}\\
 + 1.10  \ln{\left(\frac{a}{\epsilon_0} \zeta^{-3/2}\right)}^2 - 0.412  \ln{\left(\frac{a}{\epsilon_0} \zeta^{-3/2}\right)}^3 \\
+ 0.0459  \ln{\left(\frac{a}{\epsilon_0} \zeta^{-3/2}\right)}^4\Bigg].
\label{eq:rec_q}
\end{multline}
The parameters $a^* \equiv a/\epsilon_0$ and $q^* \equiv
q/\epsilon_0^2$ can then be used in Equations~(\ref{eq:es_analytic})
or (\ref{eq:el_analytic}) to find the eccentricity or ellipticity
parameter of the neutron star surface in the appropriate spacetime. As
defined in Equation~(\ref{eq:el_analytic}), the ellipticity of the
neutron star is given in Hartle-Thorne coordinates. In order to
convert to the commonly used Boyer-Lindquist coordinate system, the
following transformation can be applied:
\begin{multline}
R_{\rm BL}(R_{\rm HT}, \theta) = \\
R_{\rm HT} - \frac{a^2 \left(\frac{G M}{c^2}\right)^2}{2 R_{\rm HT}^3} \Bigg[\left(R_{\rm HT} + 2 \frac{G M}{c^2}\right)\left(R_{\rm HT} - \frac{G M}{c^2}\right)\\
- \cos^2(\theta)\left(R_{\rm HT} - 2\frac{G M}{c^2}\right)\left(R_{\rm HT} + 3 \frac{G M}{c^2}\right)\Bigg]
\label{eq:coord_transform}
\end{multline}
where $R_{\rm HT}$ is the radial coordinate in the Hartle-Thorne
coordinate system and $R_{\rm BL}$ is the radial coordinate in the
Boyer-Lindquist coordinate system (Hartle \& Thorne 1968).

\begin{figure}
\psfig{file=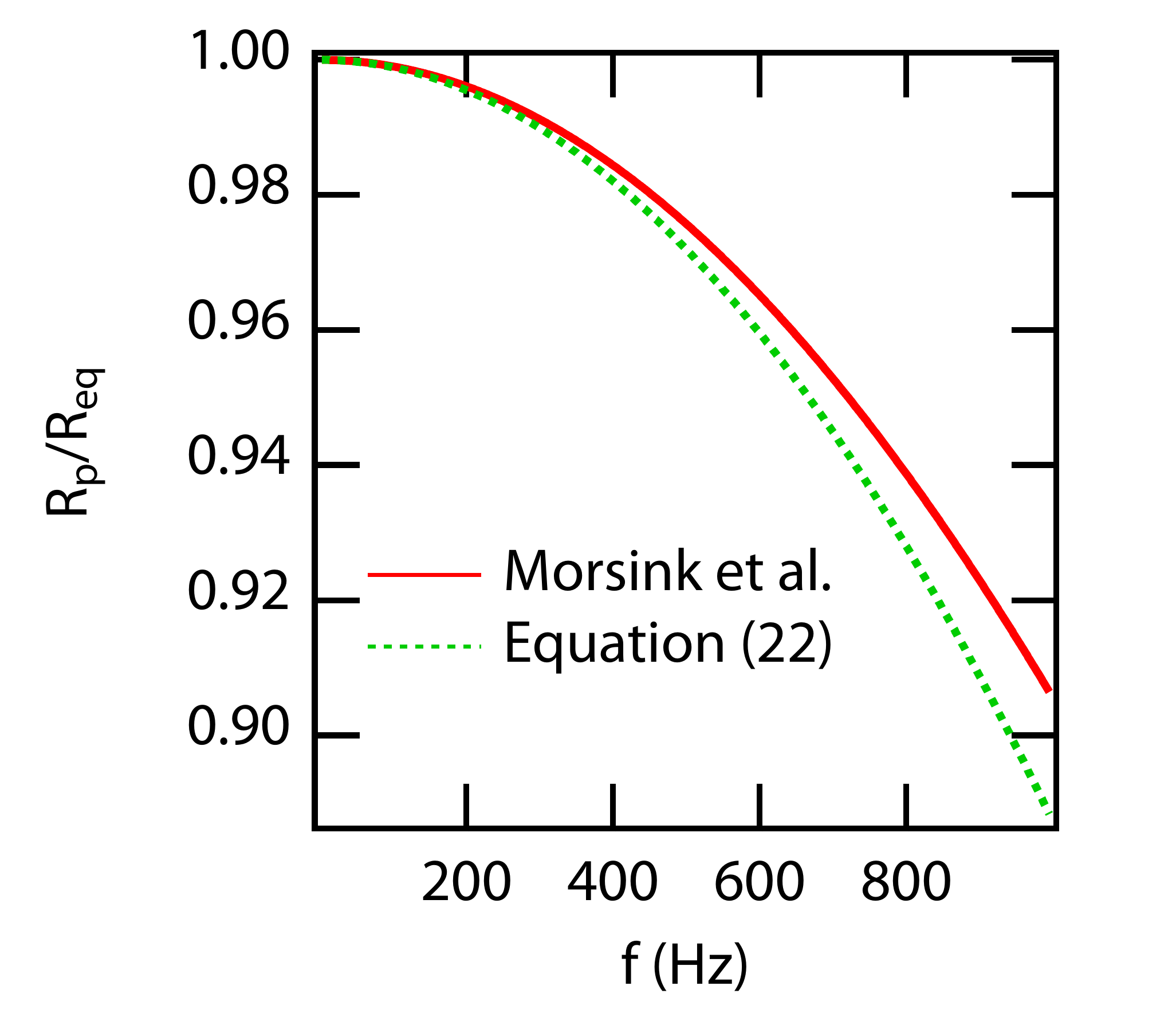, width = 3.5in}
\caption{Comparison of the ratio of polar radius to equatorial radius
  found in this paper and the ratio found by Morsink et al. (2007)}
\label{fig:Morsink_Comp}
\end{figure}

\begin{figure}
\psfig{file=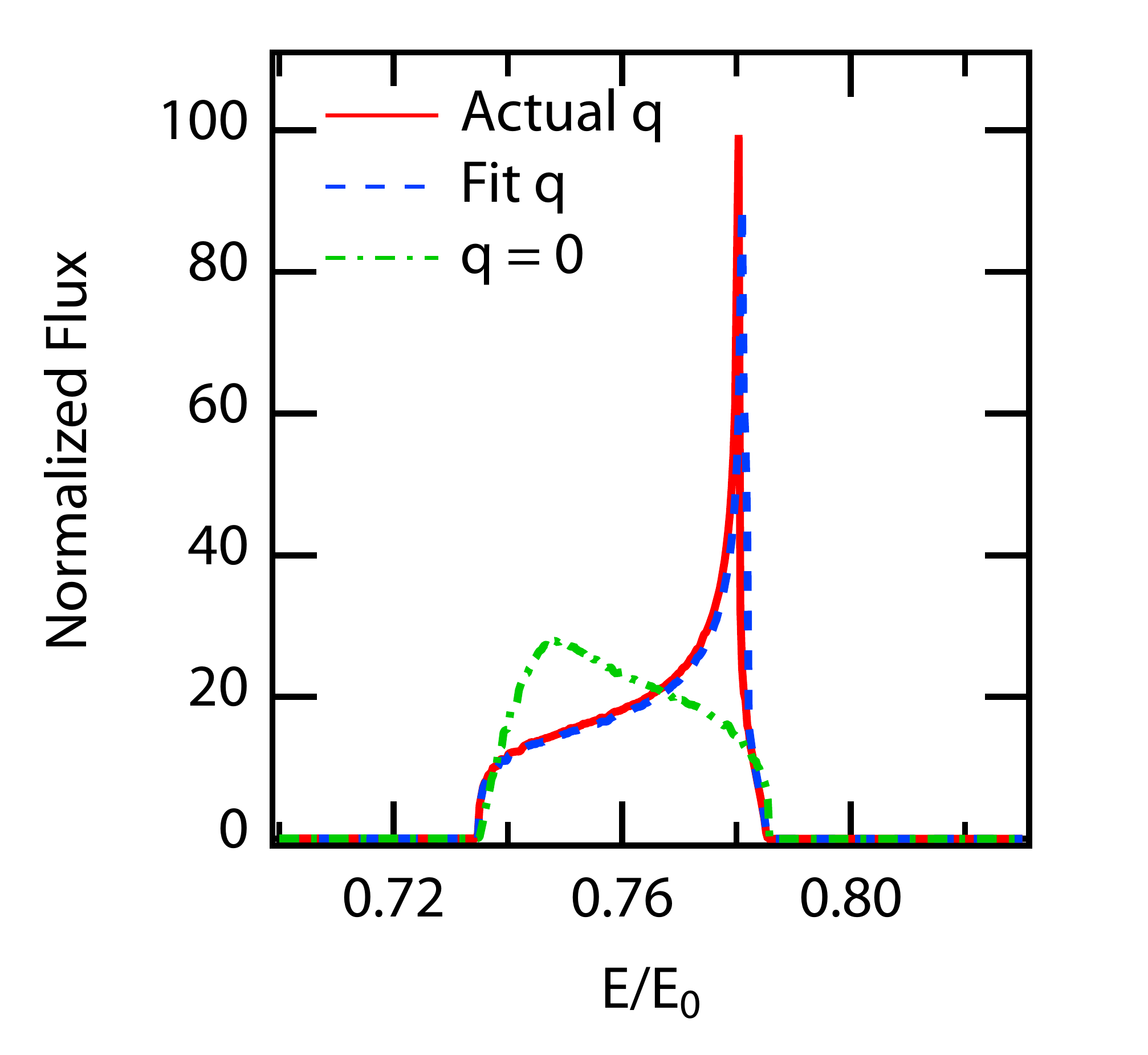, width = 3.5in}
\caption{Simulated line profiles of an emission line from a
  neutron-star surface. The solid line shows a profile calculated
  using parameters from a numerical simulation of an AP4 star with a
  mass of 1.40~$M_\odot$ and a spin frequency of 700~Hz. The dashed
  line shows a star with the same mass, spin frequency, and radius
  (10.18~km), but with the quadrupole moment, spin parameter, and
  eccentricity determined by our fits. For reference, the dash-dotted
  line shows a profile with identical parameters but with the
  quadrupole moment set to zero. The photon energy at infinity and in
  the local Lorentz frame are denoted by $E$ and $E_0$, respectively.}
\label{fig:transfer}
\end{figure}

In the context of modeling pulse profiles, Morsink et al.\ (2007)
similarly reduce the parameter space by finding an empirical
description of the oblate shape of spinning neutron stars that is
accurate for multiple equations of state. They find that compact
objects can be divided into two broad classes with different
oblateness at high spin frequencies. Normal neutron stars and hybrid
quark stars follow one relation, while color-flavor--locked stars
exhibit a different behavior. In both cases, Morsink et al.\ (2007)
find that the deviation of the stellar surface from the spherical
shape is proportional to the square of the spin frequency, with some
additional correction at fourth order in the spin.

The empirical model of Morsink et al.\ (2007) for calculating
the shape of normal neutron stars should agree with the analytic formula
we find above when compared in the same coordinate system. Morsink
et al.\ (2007) define the shape of the stellar surface in the
Schwarzschild coordinate system. Since the Boyer-Lindquist coordinate
system reduces to the Schwarzschild coordinate system in the limit of
zero spin, we use Equation~(\ref{eq:el_analytic}) to calculate the
ellipticity in Hartle-Thorne coordinates and apply the change of
coordinates described in Equation~(\ref{eq:coord_transform}). Figure~\ref{fig:Morsink_Comp}
shows the predicted ratio of the polar to the equatorial radius in 
the model of Morsink et al.\ (2007) as well as the analytic relation 
described above for a range of spin frequencies. In both models a neutron star
with a mass of 1.4 $M_\odot$ and a radius of 10~km was used.  The
deviation derived here of the empirical model of Morsink et al.\ (2007) and the analytic formula is of order 1\%
in the range of observed spins.

The neutron-star shape and quadrupole moment play an important role in
the profiles of lines that originate on neutron-star
surfaces. Baub\"ock et al.\ (2013) showed that, at low inclinations,
the quadrupole moment can cause anomalously narrow features to appear
even for neutron stars spinning at moderate rates. In order to test
whether the fits proposed in this work are precise enough to
accurately model line profiles, we compared the profile calculated
with the parameters predicted by a numerical simulation to one using
the parameters from our fits. We show the result in
Figure~\ref{fig:transfer}. For this example, we chose a model where
the fits have large residuals, especially for the quadrupole moment,
which provides the dominant contribution to the profile shape
(Baub\"ock et al 2013). Even in this case, the narrow profile is
recovered, and the difference in the resulting profiles is negligible.

\section{Conclusions}
We have demonstrated that several macroscopic parameters of spinning
neutron stars can be approximated with high accuracy using relations
that depend only on their masses, radii, and spin frequencies, but
that are practically independent of the equation of state. These fits
enable measurements of neutron-star masses and radii using X-ray
spectroscopy, timing observations of pulse profiles, and gravitational-wave
observations of neutron stars spinning at moderate frequencies.

Future detectors such as NICER, LOFT, and Advanced LIGO will soon
allow for more precise measurements of neutron-star parameters
than have been possible to date. Using these observations to
constrain the equation of state of the dense matter found in 
neutron star cores requires that the parameter space be reduced
in order to determine the mass and radius with the highest precision.
The relations demonstrated above allow this reduction of the parameter
space independent of the equation of state, making possible more precise
measurement of the equation of state of neutron-star cores.

\acknowledgements The work at the University of Arizona was supported in part by NSF grant
AST-1108753, NSF CAREER award AST-0746549, and Chandra Theory grant
TM2-13002X. E.B.'s research is supported by NSF CAREER grant
No. PHY-1055103. F.\"O.\ gratefully acknowledges support from the
Radcliffe Institute for Advanced Study at Harvard University.


\begin{thebibliography}{}
\bibitem[Andersson \& Kokkotas (1998)]{ak98} Andersson, N.\ \& Kokkotas, K.\ 1998, MNRAS, 299, 1059
\bibitem[Antoniadis et al.\ (2013)]{ant13} Antoniadis, J.\ et al. 2013,  Science, 340, 448
\bibitem[Baub\"ock et al.\ (2013)]{bau13} Baub\"ock, M., Psaltis, D., \& \"Ozel, F.\ 2013, \apj, 766, 87
\bibitem[Baub\"ock et al.\ (2012)]{bau12} Baub\"ock, M., Psaltis, D., \"Ozel, F., \& Johannsen, T.\ 2012, \apj, 753, 175
\bibitem[Berti \& Stergioulas(2004)]{bs04} Berti, E. \& Stergioulas, N.\ 2004, \mnras, 350, 1416 
\bibitem[Berti et al. (2005)]{ber05} Berti, E., White, F., Maniopoulou, A. \& Bruni, M. 2005, MNRAS, 358, 923
\bibitem[Bogdanov et al. (2008)]{bog08} Bogdanov, S., Grindlay, J., \& Rybicki, G. 2008, \apj, 689, 407
\bibitem[Brown et al.(2012)]{bhl12} Brown, D.~A., Harry, I., 
Lundgren, A., \& Nitz, A.~H.\ 2012, \prd, 86, 084017 
\bibitem[Burgay et al.(2003)]{bur03} Burgay, M., D'Amico, N., Possenti, A., et al.\ 2003, \nat, 426, 531 
\bibitem[Cook, Shapiro \& Teukolsky (1994)]{coo94} Cook, G.~B., Shapiro, S.~L. \& Teukolsky, S.~A. 1994, \apj, 424, 823
\bibitem[Demorest et al.\ (2010)]{dem10} Demorest, P., Pennucci, T., Ransom, S., Roberts, M., \& Hessels, J. 2010, Nature, 467, 7319, 1081
\bibitem[Ferdman et al.(2008)]{fer08} Ferdman, R.~D., Stairs, I.~H., Kramer, M., et al.\ 2008, 40 Years of Pulsars: Millisecond Pulsars, Magnetars and More, 983, 474 
\bibitem[Galloway et al.\ (2008)]{gal08} Galloway, D., Muno, M., Hartman, J., Psaltis, D., \& Chakrabarty D. 2008, \apj, 179, 360
\bibitem[Guillot et al.\ (2011)]{gul11} Guillot, S., Rutledge, R., \& Brown, E. 2011, \apj 732, 88
\bibitem[Hartle (1967)]{har67} Hartle, J. 1967, \apj, 150, 1005
\bibitem[Hartle \& Thorne (1968)]{ht68} Hartle, J. \& Thorne, K. 1968, \apj, 153, 807
\bibitem[Heinke et al.\ (2006)]{hei06} Heinke, C., Rybicki, G., Narayan, R., Grindlay, J. 2006, \apj, 644, 1090 
\bibitem[Hessels et al.(2006)]{hes06} Hessels, J.~W.~T., Ransom, S.~M., Stairs, I.~H., et al.\ 2006, Science, 311, 1901 
\bibitem[Lattimer \& Prakash (2001)]{lp01} Lattimer, J., \& Prakash, M. 2001, \apj, 550, 426
\bibitem[Laarakkers \& Poisson (1999)]{lp99} Laarakkers, W. \& Poisson, E. 1999, \apj, 512, 282
\bibitem[Lo \& Lin (2011)]{ll2011} Lo, K. \& Lin, L., 2011, \apj, 728, 12
\bibitem[Lorimer(2001)]{lor01} Lorimer, D.~R.\ 2001, Living Reviews in Relativity, 4, 5 
\bibitem[Mandel \& O'Shaughnessy(2010)]{mos10} Mandel, I., \& O'Shaughnessy, R.\ 2010, Classical and Quantum Gravity, 27, 114007 
\bibitem[Morsink et al. (2007)]{mor07} Morsink, S.~M., Leahy, D.~A., Cadeau, C., \& Braga, J. 2007, \apj, 663, 1244
\bibitem[\"Ozel (2013)]{oz13} \"Ozel, F. 2013, Rep.\ Prog.\ Phys., 76, 016901
\bibitem[\"Ozel et al.\ (2009)]{oz09} \"Ozel, F., G\"uver, T., \& Psaltis, D. 2009, \apj, 693, 1775
\bibitem[\"Ozel et al.(2012)]{2012ApJ...757...55O} {\"O}zel, F., Psaltis, 
D., Narayan, R., \& Santos Villarreal, A.\ 2012, \apj, 757, 55 
\bibitem[Pappas \& Apostolatos (2012)]{pa12} Pappas, G. \& Apostolatos, T. 2012, \prl, 108, 231104
\bibitem[Poisson(1998)]{po98} Poisson, E.\ 1998, \prd, 57, 5287 
\bibitem[Psaltis \& \"Ozel (2013)]{po13} Psaltis, D., \& \"Ozel, F.\ 2013,
\apj, submitted, arXiv:1305.6615
\bibitem[Racine(2008)]{ra08} Racine, {\'E}.\ 2008, \prd, 78, 044021 
\bibitem[Ravenhall \& Pethick (1994)]{rp94} Ravenhall, D., \& Pethick, C. 1994, \apj, 424, 846
\bibitem[Urbanec et al.\ (2013)]{urb13} Urbanec, M., Miller, J., \& Stuchl\'ik, Z. 2013, MNRAS, 433, 1903
\bibitem[Webb \& Barret (2007)]{wb07} Webb, N., \& Barret, D. 2007, \apj, 671, 727
\bibitem[Yagi \& Yunes (2013)]{yy13a} Yagi, K., \& Yunes N. 2013a, Science, 341, 365
\bibitem[Yagi \& Yunes (2013)]{yy13b} Yagi, K., \& Yunes N. 2013b, \prd, 88, 023009

\end{thebibliography}
\end {document}